\documentclass[prb,twocolumn,superscriptaddress,showpacs]{revtex4}

\usepackage{graphicx}
\usepackage{dcolumn}
\usepackage{amsmath}
\usepackage{color}

\begin{document}
\title{Electron spin resonance in Eu based Fe pnictides}

\author{H.-A.~Krug von Nidda}
\author{S.~Kraus}
\author{S.~Schaile}
\author{E.~Dengler}
\author{N.~Pascher}
\author{M.~Hemmida}
\affiliation{Experimentalphysik V, Center for Electronic
Correlations and Magnetism, Institute for Physics, Augsburg
University, D-86135 Augsburg, Germany}
\author{M.~J.~Eom}
\author{J.~S.~Kim}
\affiliation{Departement of Physics, Pohang University of Science
and Technology, Pohang 790-784, Korea}
\author{H.~S.~Jeevan}
\author{P.~Gegenwart}
\affiliation{I. Physik. Institut, Georg-August-Universit\"{a}t
G\"{o}ttingen, D-37077 G\"{o}ttingen, Germany }
\author{J.~Deisenhofer}
\author{A.~Loidl}
\affiliation{Experimentalphysik V, Center for Electronic
Correlations and Magnetism, Institute for Physics, Augsburg
University, D-86135 Augsburg, Germany}

\date{\today}

\begin{abstract}
The phase diagrams of EuFe$_{2-x}$Co$_x$As$_2$ $(0 \leq x \leq 0.4)$ and EuFe$_2$As$_{2-y}$P$_y$ $(0 \leq y \leq 0.43)$ are investigated by Eu$^{2+}$ electron spin resonance (ESR) in single crystals. From the temperature dependence of the linewidth $\Delta H(T)$ of the exchange narrowed ESR line the spin-density wave (SDW) $(T < T_{\rm SDW})$ and the normal metallic regime $(T > T_{\rm SDW})$ are clearly distinguished. At $T > T_{\rm SDW}$ the isotropic linear increase of the linewidth is driven by the Korringa relaxation which measures the conduction-electron density of states at the Fermi level. For $T < T_{\rm SDW}$ the anisotropy probes the local ligand field, while the coupling to the conduction electrons disappears. With increasing substitution $x$ or $y$ the transition temperature $T_{\rm SDW}$ decreases linearly accompanied by a linear decrease of the Korringa-relaxation rate from 8~Oe/K at $x=y=0$ down to 3~Oe/K at the onset of superconductivity at $x \approx 0.2$ or at $y \approx 0.3$, above which it remains nearly constant. Comparative ESR measurements on single crystals of the Eu diluted SDW compound Eu$_{0.2}$Sr$_{0.8}$Fe$_2$As$_2$ and superconducting (SC) Eu$_{0.22}$Sr$_{0.78}$Fe$_{1.72}$Co$_{0.28}$As$_2$ corroborate the leading influence of the ligand field on the Eu$^{2+}$ spin relaxation in the SDW regime as well as the Korringa relaxation in the normal metallic regime. Like in Eu$_{0.5}$K$_{0.5}$Fe$_2$As$_2$ a coherence peak is not detected in the latter compound at $T_{\rm c}=21$~K, which is in agreement with the expected complex anisotropic SC gap structure.

\end{abstract}



\maketitle

\section{Introduction}

Iron-based pnictides and chalcogenides became a central topic of modern solid-state physics when in 2008 superconductivity was discovered in LaFeAsO$_{1-x}$F$_x$ with transition temperatures as high as $T_{\rm c} = 26$~K,\cite{Kamihara08} as well as in SmFeAsO$_{1-x}$F$_x$ ($T_{\rm c} = 55$~K),\cite{Chen08,Ren2008} in Ba$_{1-x}$K$_x$Fe$_2$As$_2$ ($T_{\rm c} = 38$~K),\cite{Rotter08a,Kant2010} and in $\alpha$-FeSe$_{1-x}$ ($T_{\rm c} = 8$~K).\cite{Hsu08} Apparently the superconductivity emerges from the FeAs or FeSe layers which are the building blocks of the corresponding quasi two-dimensional crystal structures suggesting the analogy to the cuprate-based high-$T_{\rm c}$ superconductors. Similar to the cuprates, the mother compounds $(x=0)$ exhibit an antiferromagnetic ground state - here usually a spin-density wave (SDW) - which under doping $(x>0)$ becomes gradually suppressed and gives rise to a superconducting dome. However, in the iron-based superconductors the parent phases usually are metallic. Early reviews on the enormous research activities to understand the origin of the superconductivity and its interplay with the magnetism of iron in these compounds are given by Johnston \cite{Johnston2010} and Stewart.\cite{Stewart2011}

The ternary \textit{A}Fe$_2$As$_2$ (122)-systems (with \textit{A} = Ba, Sr, Ca, Eu), which crystallize in the tetragonal ThCr$_2$Si$_2$ structure, can be driven from the SDW ground state into superconductivity by chemical substitution on each of the lattice sites, e.g., substituting the $A$-site ions by K,\cite{Rotter08a,Jeevan08b} or Fe by
Co,\cite{Sefat08,Leithe-Jasper08} as well as As by P.\cite{Ren2009,Jeevan2011} Special attention is devoted to EuFe$_2$As$_2$ showing the highest reported SDW transition temperature $T_{\rm{SDW}} = 190$~K in the pnictides, while antiferromagnetic order of the europium spins is observed only below $T_{\rm N}=19$~K. \cite{Raffius93,Jeevan08a,Wu09} In this system it is possible to study the interplay between the localized Eu$^{2+}$ moments and the itinerant magnetism of the FeAs layers and its influence on superconductivity under application of hydrostatic pressure or doping.

The antiferromagnetic structure of EuFe$_2$As$_2$ was investigated by means of magnetic resonant x-ray scattering and neutron scattering.\cite{Herrero2009,Xiao2009,Koo2010} The SDW transition, which is accompanied by an orthorhombic structural distortion, was found to be weakly of first order, while the ordering of the Eu moments was characterized as second-order phase transition. Ferromagnetic Eu$^{2+}$ layers are coupled antiferromagnetically along the $c$ axis with the europium moments aligned along the $a$ axis. The iron moments are aligned along the $a$ axis as well, but directed antiparallel on neighboring sites along $a$ and $c$.
Optical spectroscopy probed the opening of the SDW gap,\cite{Wu09} which even turned out to consist of two gaps related to different electronic subsystems.\cite{Moon09} The complex electronic structure is further elucidated by angular resolved photo-electron spectroscopy (ARPES) which reveals droplet-like Fermi surfaces in the SDW state of the iron system.\cite{deJong2010} Nevertheless, there have been no detectable changes across the antiferromagnetic ordering of the Eu spins, indicating only a weak coupling between both systems.\cite{Zhou2010}
External magnetic fields reorient the Eu spins into a ferromagnetic phase already below 2~T,\cite{Jiang2009} but do not affect the iron moments at least up to highest applied fields of 55~T.\cite{Tokunaga2010} Magneto-transport measurements show that electron scattering due to the Eu$^{2+}$ local moments plays only a minor role for the electronic transport properties of EuFe$_2$As$_2$.\cite{Terashima2010a} Both observations again are in accordance with weak coupling between Fe spins and Eu spins.

At hydrostatic pressures above 20~kbar EuFe$_2$As$_2$ exhibits superconductivity below $T_{\rm c}=29.5$~K.\cite{Miclea09,Terashima2009,Uhoya2010} The maximum upper critical fields $B_{\rm c2}(T=0) \sim 25$~T are significantly reduced as compared to related systems with non magnetic ions in the $A$ site due to the pair-breaking influence of the europium spins.\cite{Kurita2011a} Detailed investigations of the phase diagram reveal the suppression of the iron SDW phase between 25 and 27 kbar, superconductivity for $24 < p < 31$~kbar,\cite{Kurita2011b} a change of antiferromagnetic Eu order into ferromagnetic order at about 60~kbar and gradual suppression of the ferromagnetic phase above 100~kbar indicating valence fluctuations of Eu$^{2+}$ to non magnetic Eu$^{3+}$.\cite{Matsubayashi2011} Concomitantly, at low temperatures the structural transition from tetragonal into collapsed tetragonal is observed at about 110~kbar.\cite{Uhoya2011}

A similar evolution of the phase diagram is observed in EuFe$_2$As$_{2-x}$P$_x$ by substitution of isovalent phosphorous on the arsenic site with superconductivity for $0.32 < x < 0.42$.\cite{Jeevan2011} Magnetization and resistivity,\cite{Cao2011} magnetic Compton scattering\cite{Ahmed2010} and M\"ossbauer studies,\cite{Nowik2011a} indicate the coexistence and competition of superconductivity and ferromagnetism, which sets in already on the right wing of the superconducting dome.\cite{Zapf2011}
Moreover, optical investigations reveal BCS-type s-wave superconductivity without nodes in EuFe$_2$As$_{1.64}$P$_{0.36}$.\cite{Wu2011} For higher P concentration in EuFe$_2$As$_{1.4}$P$_{0.6}$, resonant x-ray scattering indicates a valence change of europium like in pure EuFe$_2$As$_2$ under high pressure.\cite{Sun2010}

The electron doping by substitution of Fe by Co revealed a comparable behavior,\cite{Zheng09} but the superconducting transition remains incomplete in the resistivity and dissimilarities of the electronic structures with respect to P substitution were pointed out based on ARPES experiments in EuFe$_2$As$_{2-x}$P$_x$.\cite{Tirupathaia2011} Magnetization, magnetic torque, and nuclear magnetic resonance measurements on EuFe$_{2-x}$Co$_x$As$_2$ suggest the coupling of Fe and Eu sublattices to be dependent on Co doping.\cite{Guguchia2011a,Guguchia2011b} M\"ossbauer spectroscopy indicates that the SDW survives in the region of superconductivity\cite{Blachowski2011} and the antiferromagnetic order of the Eu spins changes into a tilted helical structure with increasing Co concentration.\cite{Nowik2011b}

Moreover, suppression of the SDW transition and emergence of ferromagnetic ordering of the Eu$^{2+}$ moments was found in EuFe$_{2-x}$Ni$_x$As$_2$, but without any superconducting phase.\cite{Ren2009a} On the other hand, dilution of the Eu system is in favor of superconductivity as observed, e.g., in Eu$_{0.5}$K$_{0.5}$Fe$_2$As$_2$ with $T_{\rm c} = 32$~K,~\cite{Jeevan08b} in Eu$_{0.7}$Na$_{0.3}$Fe$_2$As$_2$ with $T_{\rm c}= 34.7$~K,\cite{Qi2008} and under hydrostatic pressure in Eu$_{0.5}$Ca$_{0.5}$Fe$_2$As$_2$ with $T_{\rm c} \approx 20$~K above 12.7 kbar.\cite{Mitsuda2010}

Electron spin resonance (ESR) at the Eu$^{2+}$ spins turned out to be very sensitive to the local electronic density of states. Due to its half-filled $4f$ shell, Eu$^{2+}$ exhibits a pure S state with spin $S=7/2$. Because of the zero orbital momentum the direct relaxation to the lattice practically vanishes and, therefore, the paramagnetic resonance line can be well resolved up to room temperature. In single crystalline EuFe$_2$As$_2$ previous ESR investigations revealed a drastic change in the spin relaxation of the europium system from a Korringa-type linear increase of the linewidth with temperature typical for metals above $T_{\rm{SDW}}$ to an anisotropic linewidth approaching an asymptotic high-temperature value as characteristic for an insulator  in the SDW state.\cite{Dengler10} Although the resistivity is even smaller below $T_{\rm{SDW}}$ than above, the scattering of the conduction electrons at the Eu spins, which provides the Korringa relaxation, is strongly suppressed in the SDW phase, which was interpreted in the sense that the conduction electrons become strictly confined to the FeAs layers. First ESR investigations of the Co substitutional series at temperatures above 77~K indicate a linear relation between the Korringa slope at high temperatures and both the SDW transition temperature $T_{\rm SDW}$ and the superconducting transition temperature $T_{\rm c}$.\cite{Ying10} In Eu$_{0.5}$K$_{0.5}$Fe$_2$As$_2$, where the SDW is completely suppressed by hole doping and superconductivity is found below $T_{\rm c} = 32$~K,~\cite{Jeevan08b} the ESR signal of the Eu$^{2+}$ spins gives
direct access to the superconducting state:\cite{Pascher10} above $T_{\rm c}$ a normal Fermi-liquid behavior was identified from the Korringa-law of the ESR spin-lattice relaxation rate $1/T_1^{\rm ESR}$. Below $T_{\rm c}$ the spin-lattice relaxation rate followed a super-linear law $1/T_1^{\rm ESR}\propto T^{1.5}$. At the same time no Hebel-Slichter peak was observed, ruling out a simple isotropic BCS scenario in this compound. In the present work, we report a comprehensive study of the influence of cobalt and phosphorus substitution as well as dilution of europium on the ESR properties of EuFe$_2$As$_2$ single crystals in the temperature range $4.2 \leq T \leq 300$~K.

\section{Experimental details}

Single crystals of EuFe$_{2-x}$Co$_{x}$As$_2$ $(0 \leq x \leq 0.4)$ and EuFe$_{2}$As$_{2-y}$P$_{y}$ $(0 \leq y \leq 0.43)$ -- as well as two single crystals ($x=0$ and 0.28) with about 80\% Sr on the Eu site -- were prepared by flux technique and Bridgeman method as described in  Refs. \onlinecite{Dengler10} and \onlinecite{Jeevan2011}, respectively.

Most of the ESR measurements were performed in a \textit{Bruker} ELEXSYS E500 CW-spectrometer
at X-band frequencies ($\nu \approx 9.36$~GHz) in the temperature region $4.2 < T<
300$~K using an \textit{Oxford Instruments} ESR 900 continuous He gas-flow cryostat.
ESR detects the power $P$ absorbed by the sample from the
transverse magnetic microwave field as a function of the static
magnetic field $H$. The signal-to-noise ratio of the spectra is
improved by recording the derivative $dP/dH$ using lock-in technique
with field modulation. The single crystals were fixed in Suprasil (\textit{Heraeus}) quartz tubes by paraffin with the $c$-axis parallel or perpendicular to the axis of the quartz tube. A \textit{Bruker} programmable goniometer driven by a step motor allowed for orientation dependent measurements with a precision better than 1 degree.

For $x=0.3$ comparative ESR experiments have been performed at Q-band frequency ($\nu \approx 34.1$~GHz) also in Augsburg as well as at L-band frequency ($\nu \approx 1.09$~GHz) at the Bruker ELEXSYS spectrometer in Dresden.

\section{Experimental Results and Discussion}

\subsection{Co substitution on the Fe site}

\begin{figure}
\centering
\includegraphics[width=75mm,clip]{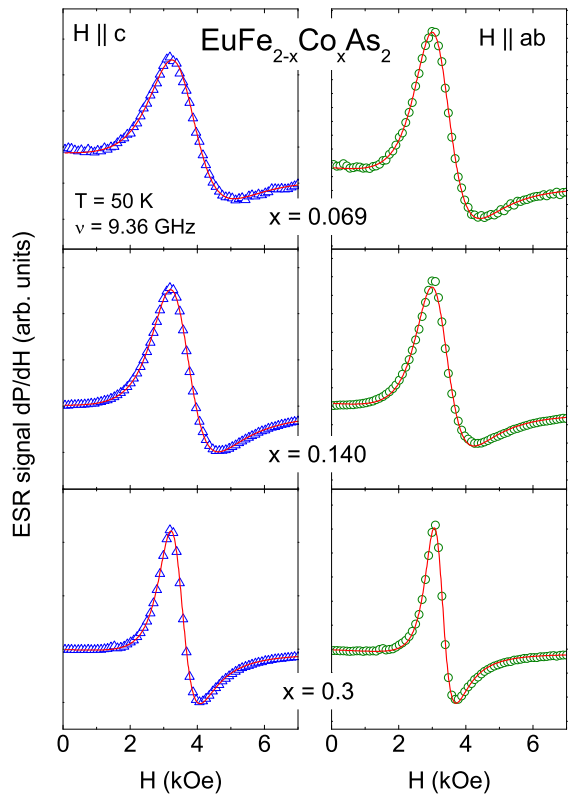}
\vspace{2mm} \caption[]{\label{spectraCobalt} (Color online) ESR spectra
obtained on single crystals of EuFe$_{2-x}$Co$_{x}$As$_2$ with $x=0.069$ (upper), 0.140 (middle), and 0.3 (lower frames) at $T = 50$~K for the magnetic field applied parallel (left column) or perpendicular (right column) to the tetragonal $c$ axis. Red solid lines indicate the fit by a Dyson shape.}
\end{figure}

Figure \ref{spectraCobalt} exemplarily shows ESR spectra of three EuFe$_{2-x}$Co$_{x}$As$_2$ single crystals with different Co concentrations $x$ in the paramagnetic regime at a temperature of $T=50$~K for the magnetic field $H$ applied both parallel and perpendicular to the crystallographic $c$ axis.
Like in the mother compound \cite{Dengler10} (see also upper frames of Fig.~\ref{spectraPhosphor}) all spectra consist of a single exchange-narrowed resonance line which narrows on increasing Co concentration $x$. The resonance line
is well described by a Dyson shape,\cite{Barnes1981} i.e. a Lorentz
line at resonance field $H_{\rm res}$ with half width at half
maximum $\Delta H$ and a contribution of dispersion to the
absorption given by the $(D/A)$ ratio $0 \leq D/A \leq 1$. The dispersion results in an
asymmetry typical for metals, where due to the conductivity the skin effect drives electric
and magnetic components of the microwave field out of phase in the sample. The
$D/A$ ratio depends on the sample size, geometry, and skin depth. If the skin depth is small compared
to the sample size, $D/A$ approaches 1, in the reverse case $D/A$ vanishes like in insulators. As the linewidth $\Delta H$ is
of the same order of magnitude as the resonance field $H_{\rm res}$, the counter resonance
at $-H_{\rm res}$ was included in the fitting process as well.\cite{Joshi2004}

\begin{figure}
\centering
\includegraphics[width=80mm,clip]{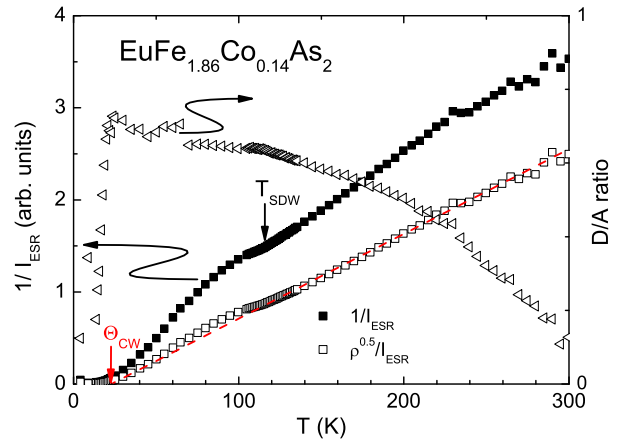}
\vspace{2mm} \caption[]{\label{intensity} (Color online) Dispersion-to-absorption ratio (right ordinate) and inverse ESR intensity (left ordinate) before (closed squares) and after (open squares) skin-depth correction for a single crystal EuFe$_{1.86}$Co$_{0.14}$As$_2$. The red dashed line indicates a Curie-Weiss like behavior.}
\end{figure}

Figure~\ref{intensity} exemplarily illustrates the typical temperature dependence for $D/A$ ratio and inverse ESR intensity as obtained for the cobalt concentration $x=0.14$. Similar to pure EuFe$_2$As$_2$ \cite{Dengler10} the $D/A$ ratio increases on decreasing temperature from about 0.1 at room temperature up to 0.7 below 100~K due to the decreasing electrical resistivity. A significant drop marks the onset of magnetic order, which is accompanied by a change in signal shape due to strong demagnetization effects. These are strongly dependent on details of sample shape and surface and, therefore, will not be considered further. The inverse intensity $1/I_{\rm ESR}$ monotonously increases with increasing temperature and -- after correction by the skin depth $\propto \sqrt{\rho/\nu}$ with the resistivity ($\rho$) data \cite{Ying10} and microwave frequency $\nu = 9.36$~GHz -- nicely resembles the expected Curie-Weiss law $1/I_{\rm ESR} \propto (T-\Theta_{\rm CW})$ with $\Theta_{\rm CW} \approx 22$~K, except of a slight shift below the SDW transition at $T_{\rm SDW} \approx 120$~K.

\begin{figure}[t]
\centering
\includegraphics[width=70mm]{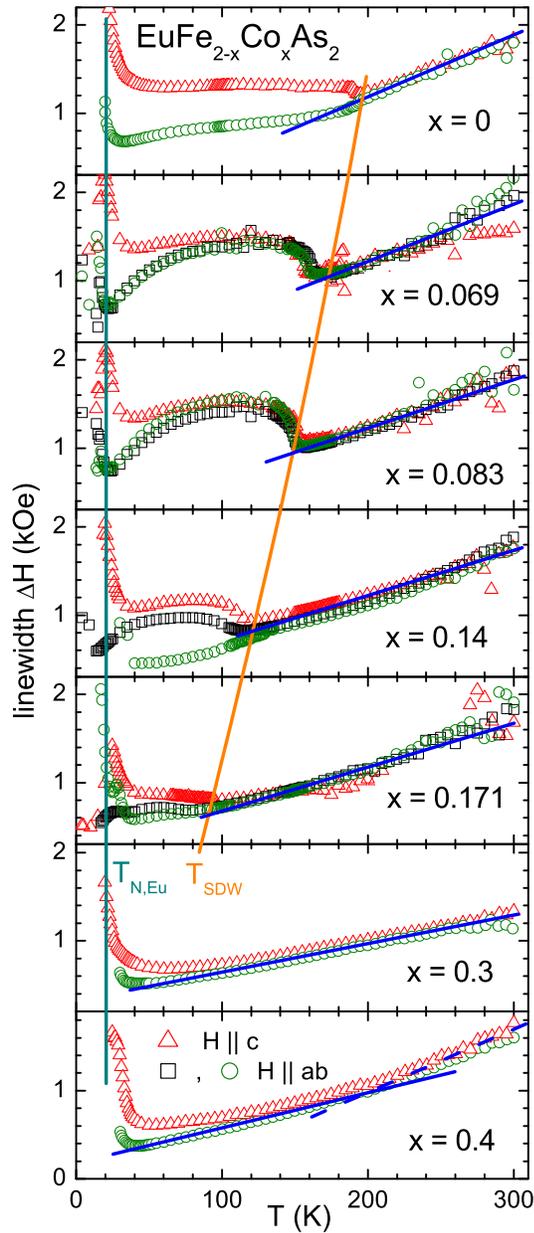}
\vspace{2mm} \caption[]{\label{dh} (Color online) Temperature
dependence of the ESR linewidth $\Delta H$ in EuFe$_{2-x}$Co$_{x}$As$_2$ for $0 \leq x \leq 0.3$ for the magnetic field aligned along the principal axes. The concentration dependence of the SDW transition ($T_{\rm SDW}$) and of the magnetic ordering transition of the Eu spins ($T_{\rm N,Eu}$) is illustrated by solid lines. The straight solid lines in each graph
indicate the linear Korringa law.}
\end{figure}

\begin{figure}[t]
\centering
\includegraphics[width=70mm]{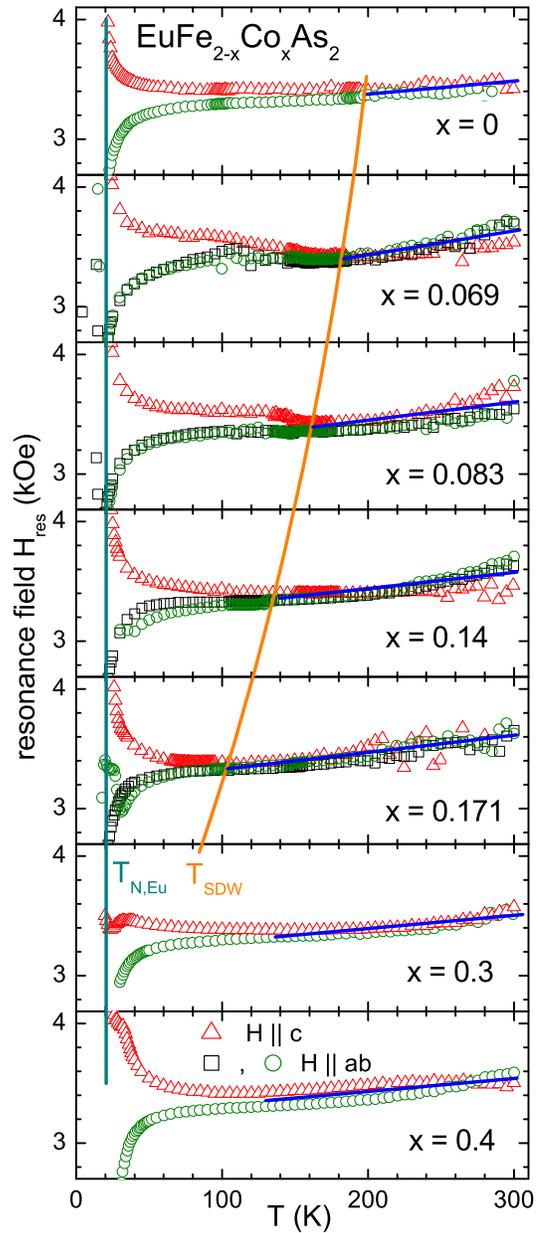}
\vspace{2mm} \caption[]{\label{hres} (Color online) Temperature
dependence of the resonance field $H_{\rm res}$ in EuFe$_{2-x}$Co$_{x}$As$_2$ for $0 \leq x \leq 0.3$ for the magnetic field aligned along the principal axes. The concentration dependence of the SDW transition ($T_{\rm SDW}$) and of the magnetic ordering transition of the Eu spins ($T_{\rm N,Eu}$) is illustrated by solid lines. The blue solid lines in each graph are guides for the eye.}
\end{figure}

The most important information is obtained from the temperature dependence of the linewidth and the resonance field depicted in Figs.~\ref{dh} and \ref{hres} for the magnetic field aligned along the three principal crystal axes. For completeness we added the data of pure EuFe$_2$As$_2$. Up to $x=0.171$ one clearly recognizes the SDW transition at $T_{\rm SDW}$, which separates the usual metallic Korringa-relaxation regime at high temperatures from the insulator-like relaxation regime at low temperatures. Moreover, the resonance field significantly shifts on approaching magnetic order of the Eu spins at $T_{\rm N}$ due to the increasing demagnetization fields, at the same time the linewidth strongly increases because of critical magnetic fluctuations close to the phase transition.

In contrast to the isotropy of the parameters above the SDW transition, below $T_{\rm SDW}$ both linewidth and resonance field develop a pronounced spatial anisotropy. In the SDW state only the dipolar interaction between the Eu spins and the crystal-electric field of the ligands determine the spin relaxation like in an insulator. While for $x=0$ the anisotropy shows up directly on crossing $T_{\rm SDW}$ down to lower temperatures, one only observes a strong increase of the linewidth for $x=0.069$ and $x=0.083$ followed by a rather temperature independent regime. Only at lower temperatures ($T<T_{\rm SDW}/2$) the tetragonal symmetry shows up. This is probably a consequence of the disorder induced by the Co substitution, which randomly disturbs the local crystal field and its symmetry axis leading to an averaged broadening. Note, however, that for $x \leq 0.1$ the absolute value of the linewidth in this intermediate regime approximates the value of the plateau for $H \parallel c$ at $x=0$, indicating a comparable magnitude of the uniaxial zero-field splitting parameter $D$, which was determined as $D \approx 5.5$~GHz in EuFe$_2$As$_2$.\cite{Dengler10}

\begin{figure}
\centering
\includegraphics[width=80mm,clip]{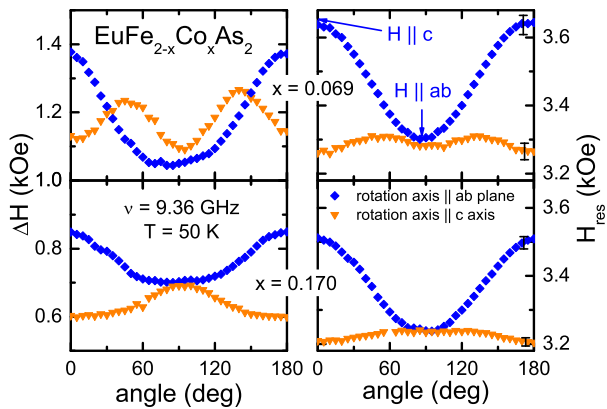}
\vspace{2mm} \caption[]{\label{angular} (Color online) Angular dependence of linewidth (left frames) and resonance field (right frames) for EuFe$_{2-x}$Co$_{x}$As$_2$ single crystals with $x=0.069$ and 0.170 at $T=50$~K.}
\end{figure}

Taking a closer look to the anisotropy in the SDW phase one recognizes the evolution of an orthorhombic contribution in contrast to the purely tetragonal symmetry in EuFe$_2$As$_2$.\cite{Dengler10} This develops on increasing Co concentration $x$ as illustrated in Figure~\ref{angular}. While for $x=0.069$ the symmetry is still tetragonal, a strong orthorhombic component shows up for $x=0.170$. Note that this development is best visible in the linewidth data, while the symmetry of the resonance field is basically determined by demagnetization: for the platelet-shaped samples under investigation the resonance shifts to higher fields in case of a magnetic field applied perpendicular to the plane, but to lower fields, if it is applied within the plane.\cite{Kittel1948} Indeed, the anisotropy of the linewidth is expected to exhibit an orthorhombic symmetry in the whole SDW regime due to the structural transition which accompanies the formation of the SDW. However, the single crystals are usually heavily twinned in the $ab$ plane, as has been revealed by Tanatar \textit{et al.}\cite{Tanatar2009} using polarized light microscopy and spatially resolved high-energy synchrotron x-ray diffraction. Nevertheless, the in-plane anisotropy of the resistivity was investigated by Chu \textit{et al.}\cite{Chu2010} in single crystals of BaFe$_{2-x}$Co$_x$As$_2$ which were detwinned by means of uniaxial stress. In the SDW phase a significant anisotropy shows up with the resistivity along the shorter $b$ axis being larger than along the $a$ axis. The in-plane anisotropy is only weak in the pure system $(x=0)$ but reaches a maximum value of about 2 for compositions close to the onset of the superconducting dome. Our ESR data which probe the anisotropy on a local atomic scale are in accordance with this finding regarding the data of the at least partially untwinned crystals with $x=0.140$ and $x=0.171$. However, ESR measurements under uniaxial pressure have to be performed in future to systematically detwin single crystals and, thus, to derive more quantitative information on the true anisotropy.

\begin{figure}
\centering
\includegraphics[width=80mm]{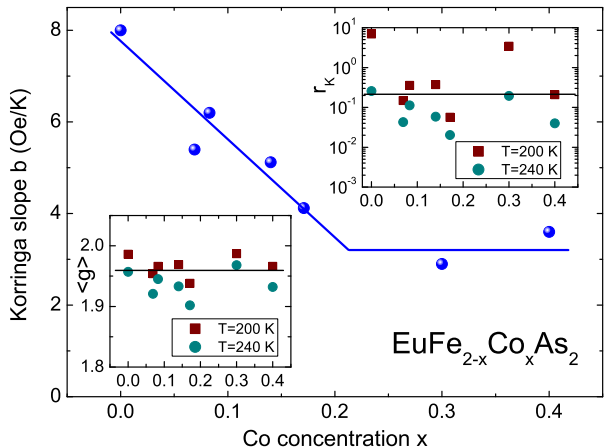}
\vspace{2mm} \caption[]{\label{Korringa} (Color online) Concentration dependence of Korringa slope $b$ (main frame) in the normal metallic regime, averaged $g$ value $\langle g \rangle$ at 200~K and 240~K (lower inset), and corresponding Korringa ratio $r_{\rm K}$ (upper inset) in EuFe$_{2-x}$Co$_{x}$As$_2$. Solid lines are drawn to guide the eye}
\end{figure}

As already observed in pure EuFe$_2$As$_2$ the linewidth and resonance field are practically independent on the orientation of the magnetic field above $T_{\rm SDW}$. Here the linewidth increases approximately linearly with temperature indicating the Korringa relaxation of the localized Eu$^{2+}$ spins via scattering of the conduction electrons\cite{Barnes1981}
\begin{equation}
\Delta H = b T = \frac{\pi k_{\rm B}}{g \mu_{\rm B}} \langle J^2(q)\rangle N^2(E_{\rm F})T
\end{equation}
where $\langle J^2(q)\rangle$ denotes the squared exchange constant between localized spins and conduction electrons averaged over the momentum transfer $q$, $N(E_{\rm F})$ is the conduction-electron density of states at the Fermi energy $E_{\rm F}$, $\mu_{\rm B}$ denotes the Bohr magneton and $k_{\rm B}$ the Boltzmann constant. With increasing Co concentration $x$ the Korringa slope $b$, which is depicted in Fig.~\ref{Korringa}, decreases from $b \approx 8$~Oe/K until the SDW is suppressed and remains approximately constant at about $b \approx 3$~Oe/K for higher $x$ values in good agreement with the results of Ying \textit{et al.},\cite{Ying10} who pointed out the empirical relation between $b$ and $T_{\rm SDW}$. Looking closer to the increase of the linewidth it turns out to be slightly stronger than linear with temperature. This effect is most pronounced for $x=0.4$, where the slope of the linewidth above 200~K is higher by a factor of about 2 as compared to temperatures below 200~K. An analogous result was obtained from $^{75}$As nuclear magnetic resonance (NMR) measurements in BaFe$_{2-x}$Co$_x$As$_2$, where in the normal metallic phase the nuclear spin-lattice relaxation rate $1/T_1$ was found to deviate from the linear temperature law in such a way that $1/(T_1T)$ \textit{vs} $T$ exhibits a positive curvature.\cite{Ning2010} The deviations from a linear temperature dependence in both ESR linewidth and nuclear spin-lattice relaxation rate are probably related to the unconventional linear increase of the conduction-electron susceptibility, as observed in iron pnictides like BaFe$_{2-x}$Co$_x$As$_2$ or LaFeAsO,\cite{Johnston2010} which is in contrast to the temperature independent Pauli susceptibility of usual metals. These findings are further supported by the temperature dependence of the resonance field.

For all Co concentrations under investigation the resonance field slightly increases with increasing temperature above $T_{\rm SDW}$. As already observed in pure EuFe$_2$As$_2$ the resulting $g$ shift is negative with respect to the insulator value $g=1.993$ and its absolute value increases with temperature. As in usual metals the $g$ shift is a local probe for the Pauli susceptibility of the conduction electrons,\cite{Barnes1981}
\begin{equation}
\Delta g = J(0)N(E_{\rm F}) \propto \chi_{\rm Pauli}
\end{equation}
in the present case such a linear increase of the $g$ value is again in accord with the linear increase of the conduction-electron susceptibility. At the same time the $g$ shift matches the linear temperature increase of the $^{75}$As Knight shift found in the NMR investigations of BaFe$_{2-x}$Co$_x$As$_2$.\cite{Ning2010}

However, it is important to mention that in the case of broad resonance lines one has to be careful with the absolute value of the resonance shift, because $D/A$ ratio and resonance field $H_{\rm res}$ are usually not fully independent parameters in such a way that an increase of $D/A$ results in a concomitant decrease of $H_{\rm res}$, i.e. increase of $g$-value. Especially for $T > 250$~K the resonance shift is accompanied by a significant drop of the $D/A$ ratio and the resulting $g$ shift $|\Delta g| > 0.1$ appears too large regarding the Korringa ratio
\begin{equation}
r_{\rm K} = \frac{g \mu_{\rm B}}{\pi k_{\rm B}} \frac{b}{(\Delta g)^2}
\end{equation}
which is expected to be $\sim 1$ in case of $s$-character and $\sim 0.2$ in case of $d$-character of the Fermi surface.\cite{Seipler1975} For $x=0$ with $b \approx 8$ and $\Delta g \approx -0.04$ (determined at 240~K sufficiently above the SDW transition) we find $r_{\rm K} \approx 0.21$ suggesting that the Eu spin basically couples to the $d$-electrons. As shown in the upper inset of Fig.~\ref{Korringa}, for all other Co concentrations the Korringa ratio determined from the $g$ value averaged over the three principal axes at two temperatures (200~K and 240~K, shown in the lower inset of the same Figure) and the average of the Korringa slope $b$ also scatters around $r_{\rm K} = 0.2$ but with large uncertainties. Thus one should not overinterpret its meaning.

\begin{figure}[t]
\centering
\includegraphics[width=80mm]{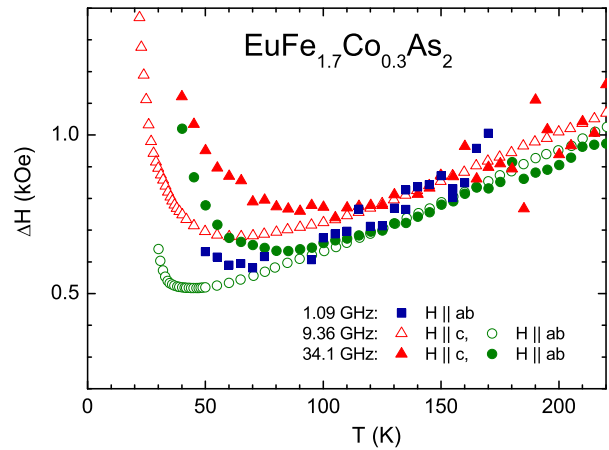}
\vspace{2mm} \caption[]{\label{LXQ} (Color online) Temperature
dependence of the linewidth in EuFe$_{1.7}$Co$_{0.3}$As$_2$ for the magnetic field aligned along the principal axes at three different microwave frequencies.}
\end{figure}

It is more important to return to the linewidth data. Recent ESR experiments by Garcia \textit{et al.}\cite{Garcia2012} performed on polycrystalline EuFe$_{2-x}$Co$_x$As$_2$ stated a significant frequency dependence of the Korringa slope in the regime $1 \leq \nu \leq 34$~GHz suggesting some kind of bottleneck scenario. We therefore investigated a single crystal with Co concentration $x=0.3$, where the linear Korringa regime extends over the broadest temperature range compared to the other concentrations, at three frequencies 1.09 (L-), 9.36 (X-), and 34.1~GHz (Q-band). As one can see in Figure~\ref{LXQ}, above 100~K the linewidth data taken at different frequencies basically coincide with each other. For X- and Q-band both for $H || c$ as well as $H \perp c$ we observe the same Korringa slope up to about 200~K. At higher temperature, the signal quality was not satisfying any more at Q-band frequency because of the smaller skin depth as compared to X-band. Below 100~K the linewidth starts to increase with decreasing temperature. This effect is stronger at Q-band frequency than at X-band due to the larger polarization of the Eu spins in the higher resonance field. L-band data could only be obtained for the orientation of smallest linewidth $H \perp c$, because the sensitivity is much weaker than in the X-band. Again the observed Korringa slope agrees with that of the higher frequencies within experimental uncertainties in contrast to the strong increase of the linewidth, which was reported for polycrystalline material.\cite{Garcia2012} Thus, our data unequivocally discard a bottleneck scenario and strongly favor the pure Korringa relaxation process. The reason is not clear at the moment, but maybe surface effects dominate in the powder samples reported in Ref.~\onlinecite{Garcia2012}. Another proof for the Korringa process will be given when considering the strontium diluted crystals below.

\subsection{P substitution on the As site}

\begin{figure}
\centering
\includegraphics[width=75mm,clip]{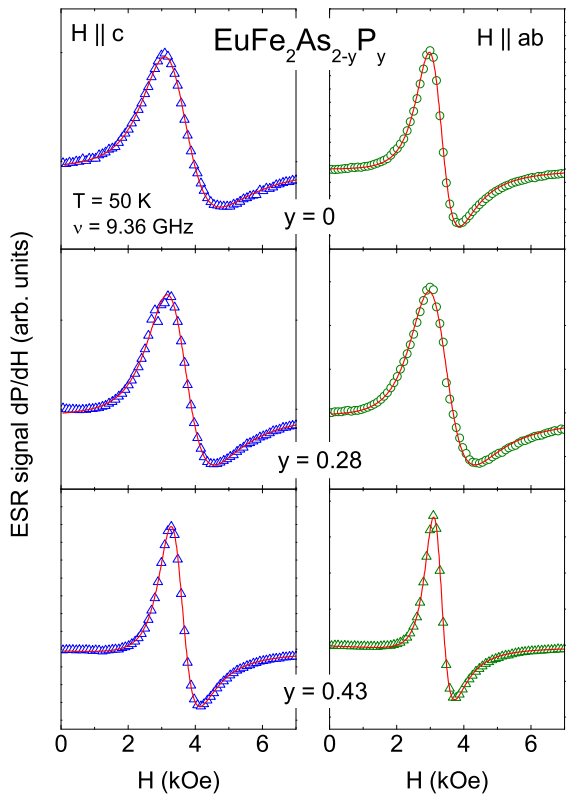}
\vspace{2mm} \caption[]{\label{spectraPhosphor} (Color online) ESR spectra
and corresponding fit curves for single crystals of EuFe$_{2}$As$_{2-y}$P$_y$ with $y=0$ (upper), 0.28 (middle), and 0.43 (lower frames) at $T = 50$~K for the magnetic field applied parallel (left column) or perpendicular (right column) to the tetragonal $c$ axis.}
\end{figure}

\begin{figure}[t]
\centering
\includegraphics[width=75mm]{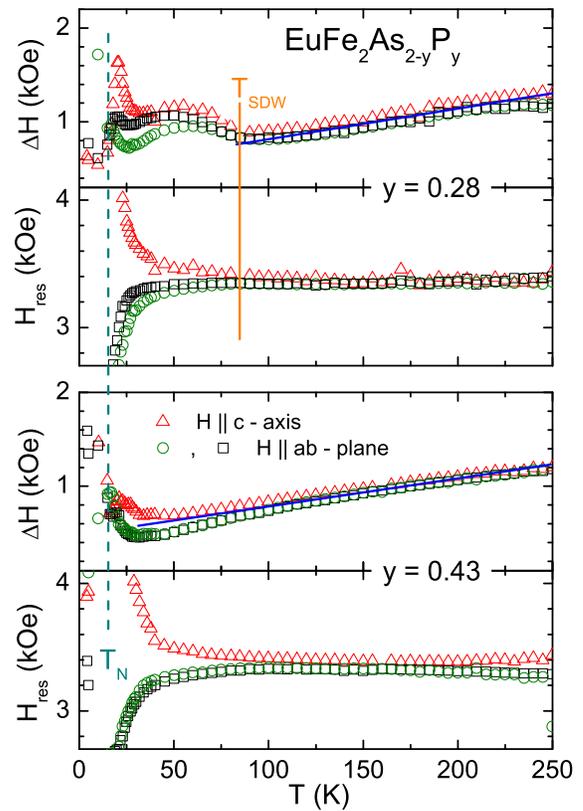}
\vspace{2mm} \caption[]{\label{PHresDH} (Color online) Temperature dependence of the ESR linewidth $\Delta H$ in EuFe$_{2}$As$_{2-y}$P$_y$ for $y=0.28$ and $y=0.43$. The SDW transition ($T_{\rm SDW}$) and the magnetic ordering transition of the Eu spins ($T_{\rm N,Eu}$) are marked by vertical lines. The blue solid lines indicate the linear Korringa law.}
\end{figure}

Before discussing the case of crystals with dilution at the Eu site, we shortly compare the case of P substitution on the As site. Figure~\ref{spectraPhosphor} shows the evolution of the ESR spectra in EuFe$_2$As$_{2-y}$P$_y$ with increasing P concentration. Again the spectra are well described in terms of a single Dyson line. The temperature dependence of linewidth and resonance field is depicted in Fig.~\ref{PHresDH}. For $y=0.28$ the characteristics of the SDW still show up below $T_{\rm SDW}\approx 80$~K. This means that in case of isoelectronic substitution (As by P) one needs about twice the concentration to suppress the SDW as compared to the electron doping case (Fe by Co). No SDW is formed for the sample with $y=0.43$. But as in the case of Co substitution the superconducting transition is masked by the onset of magnetic order of the Eu spins.

On increasing temperature an additional broad background signal becomes visible, which probably results from small amounts of residual ferromagnetic clusters $(< 1 \%)$ below the sensitivity of x-ray diffraction. This signal strongly disturbs the evaluation of the ESR spectra above 150~K and, therefore, the results have to be considered with care at those temperatures.
At least one can state that the slope of the linewidth in the usual metallic regime is reduced in a similar way as in the case of Co substitution.

Moreover it is important to note that for both P concentrations under consideration the increase of the linewidth with increasing temperature turns out to be really linear as in a usual metal. At the same time the $g$ shift from the insulator value is approximately temperature independent indicating a constant Pauli susceptibility. Indeed this is fully in accordance with $^{31}$P NMR spin-lattice relaxation rate and Knight shift in BaFe$_2$As$_{2-x}$P$_x$. The former approaches  a linear Korringa law in the normal metallic regime, while the latter resembles a temperature independent Pauli susceptibility.\cite{Nakai2010}

These results indicate that -- in contrast to the Co substitution -- P substitution drives the material rather into a normal Pauli-like metallic state.

\subsection{Sr substitution on the Eu site}

\begin{figure}
\centering
\includegraphics[width=75mm,clip]{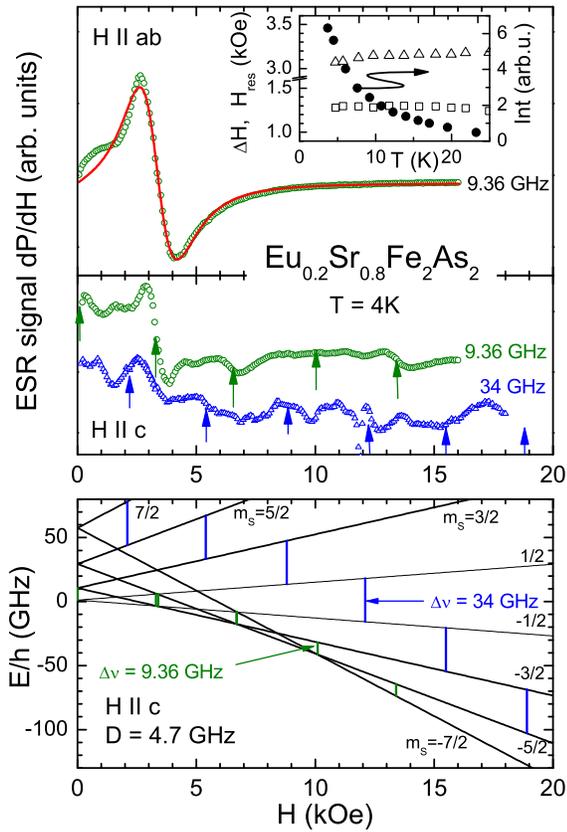}
\vspace{2mm} \caption[]{\label{Sr} (Color online) ESR spectra
and corresponding fit curves for a single crystal of Eu$_{0.2}$Sr$_{0.8}$Fe$_{2}$As$_2$ at $T = 4$~K for the magnetic field applied perpendicular (upper frame, X-band) and parallel (middle frame, X- and Q-band) to the tetragonal $c$ axis. The lower frame shows the Zeeman splitting of the ground state for the case of a zero-field splitting $D=4.7$~GHz with the corresponding magnetic dipolar transtions at X-band and Q-band frequency. Inset: temperature dependence of ESR intensity (solid spheres), resonance field (open triangles), and linewidth (open squares) for $H \perp c$.}
\end{figure}

\begin{figure}
\centering
\includegraphics[width=75mm,clip]{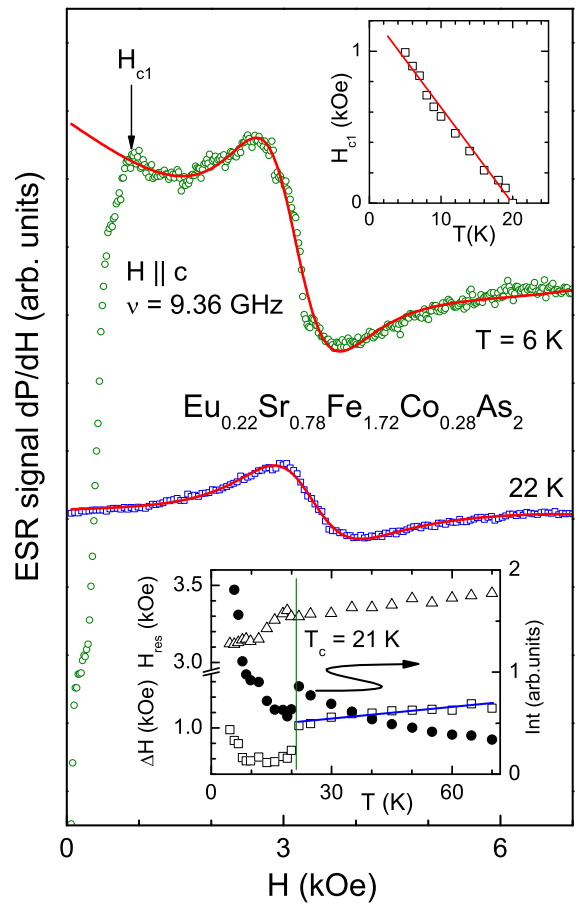}
\vspace{2mm} \caption[]{\label{SrCo} (Color online) ESR spectra
and corresponding fit curves of a single crystal of Eu$_{0.22}$Sr$_{0.78}$Fe$_{1.72}$Co$_{0.28}$As$_2$ at $T = 4$~K for the magnetic field applied parallel to the tetragonal $c$ axis. Upper Inset: lower critical field $H_{\rm c1}$ as function of temperature. Lower Inset: temperature dependence of ESR intensity (solid spheres), resonance field (open triangles), and linewidth (open squares).}
\end{figure}

To reduce the influence of the europium spins we diluted the Eu$^{2+}$ ions by substitution of about $80\%$ isovalent Sr$^{2+}$ ions both in the SDW mother compound and in a superconducting sample with $14\%$ Co on the Fe site. To start with the SDW compound, Fig.~\ref{Sr} shows spectra at $T=4$~K for the magnetic field applied parallel as well as perpendicular to the crystallographic $c$ axis. While in the latter case the spectrum appears to be reasonably exchange narrowed (upper frame), a partially resolved fine structure shows up for the former field configuration. The lower frame shows a simulation of the magnetic dipolar transitions for $H||c$ due to the purely axial zero-field splitting term $H_{\rm CF}=DS_z^2$ which modifies the Zeeman energy levels correspondingly as
\begin{equation}
E(H,m_S)=g \mu_{\rm B} H m_S + D m_S^2
\end{equation}
with $m_S = -7/2, -5/2,...,7/2$. For the calculation we used the axial parameter $D=4.7$~GHz, which simultaneously approximates the observed splitting both for X- and Q-band frequency and is in agreement with $D=5.5$~GHz obtained from the evaluation of the line-width anisotropy in EuFe$_2$As$_2$.\cite{Dengler10} As the all-over zero-field splitting $\Delta\nu_{\rm ZFS} = 12 D = 56.4$~GHz is much larger than the X-band frequency of $\nu = 9.36$~GHz, the transitions basically occur in the lower part of the diagram and partially overlap, while for the higher Q-band frequency the transitions appear already in the right order, but cannot be followed up to the highest resonance field because of the limit of the electromagnet. Deviations may occur due to disorder of the random substitution, due to weak non-zero orthorhombic contributions $E(S_y^2-S_x^2)$ to the zero-field splitting as well as from a slight misalignment $(<3^{\circ})$ of the crystal. The inset in the upper frame documents the temperature dependence of the ESR parameters obtained from the fit with a single Dyson line for $H \perp c$, which can be reasonable evaluated only up to about 25~K. At higher temperatures an impurity signal due to small amounts of ferromagnetic impurities dominates the spectrum. Besides the Curie-Weiss law of the intensity and the resonance field close to $g=2$, the nearly constant linewidth $\Delta H = 1.3$~kOe has to be mentioned. This is in good agreement with the linewidth observed in the SDW regime and also does not exhibit any linear Korringa law which again indicates the decoupling between localized Eu spins and conduction electrons in this phase.

Characteristic spectra of the superconducting compound are depicted in Fig.~\ref{SrCo} above and below the superconducting transition at $T_{\rm c}=20$~K. One observes a single exchange narrowed line which is again well described in terms of a Dyson line. In the superconducting regime it is accompanied by a broad non resonant microwave absorption due to shielding currents, which allows the determination of the lower critical field $H_{\rm c1}$ at the transition from the Meissner phase into the Shubnikov phase, where magnetic flux enters the sample. The upper inset shows the temperature dependence of $H_{c1}$ which increases approximately linearly on decreasing temperature from $T_{\rm c}$. The lower inset illustrates the temperature dependence of the ESR parameters of the paramagnetic resonance line. In the superconducting phase the broad background was empirically approximated by a second Dyson shape. As in Eu$_{0.5}$K$_{0.5}$Fe$_2$As$_2$,\cite{Pascher10} one clearly recognizes a drop in the intensity on cooling through $T_{\rm c}$ due to the shielding effect in the superconducting phase and a shift of the resonance field to $g$ values $g \geq 2$. At the same time the linewidth drops at $T_{\rm c}$ without any coherence peak, ruling out a simple isotropic BCS scenario in agreement with recent NMR results.\cite{Sarkar2012} However the exact behavior close to $T_{\rm c}$ cannot be resolved because of the strong temperature dependence of the non resonant microwave absorption at the onset of superconductivity. Nevertheless, the linear increase of the linewidth above $T_{\rm c}$ with a slope of $b=3.3$~Oe/K agrees with that observed in EuFe$_{1.7}$Co$_{0.3}$As$_2$, thus, proving the Korringa relaxation in the normal metallic phase and ruling out any bottleneck effect between the Eu spins and the conduction electrons, which would depend on the concentration of the Eu spins. Again a slight linear increase shows up in the resonance field indicating its relation to the linear increase of the conduction-electron susceptibility.

\section{Summary}

Electron spin resonance of Eu$^{2+}$ ($4f^7$, $S=7/2$) in europium based iron pnictides successfully probes the local density of states of the conduction electrons. Starting from the mother compound EuFe$_2$As$_2$, the usual metallic phase is characterized by the linear increase of the linewidth on increasing temperature (Korringa slope $b = 8$~Oe/K) due to the Korringa relaxation via the conduction electrons, while this relaxation contribution is switched off in the spin-density wave phase ($T < T_{\rm SDW}$), where the linewidth is mainly determined by the crystal-electric field of the ligands like in insulators despite of high conductivity. Substitution of cobalt for iron or of phosphorous for arsenic gradually suppresses the SDW phase and reduces the slope of the linear increase of the linewidth above $T_{\rm SDW}$ down to about $b = 3$~Oe/K. This indicates a decreasing conduction-electron density of states at the Fermi energy on increasing Co or P substitution. For further increasing substitution above the superconducting phase the Korringa slope remains approximately constant at $b \approx 3$~Oe/K. A closer inspection of the linewidth data in EuFe$_{2-x}$Co$_{x}$As$_2$ revealed a slight increase of the Korringa slope with increasing temperature. Concomitantly, the $g$ shift increases linearly with temperature. In accordance with NMR results of the related Ba compound, these observations were related to the linear temperature dependence of the conduction-electron susceptibility.

While in the Eu-concentrated compounds the superconducting properties remain masked due to the onset of magnetic order of the Eu spins close to 20~K, dilution of the europium spins by strontium substitution gives further insight: The strength of the crystal-electric field and corresponding zero-field splitting parameter $D \approx 5$~GHz determined from the anisotropy of the line width at the Eu site was corroborated by the fine-structure splitting in Eu$_{0.2}$Sr$_{0.8}$Fe$_{2}$As$_2$. The fact that the Korringa slope $b=3.3$~Oe/K observed in Eu$_{0.22}$Sr$_{0.78}$Fe$_{1.72}$Co$_{0.28}$As$_2$ approximately equals the slope of the undiluted Eu compound rules out any bottleneck effect and corroborates the direct relation between the Korringa slope and the conduction-electron density of states. This was further supported by the fact that the Korringa slope in EuFe$_{1.7}$Co$_{0.3}$As$_2$ was found to be independent on the microwave frequency. No coherence peak is observed in the temperature dependence of the linewidth close to the superconducting transition in Eu$_{0.22}$Sr$_{0.78}$Fe$_{1.72}$Co$_{0.28}$As$_2$ indicating the anisotropic structure of the superconducting energy gap. Detailed ESR investigations have to be performed on a series of single crystals with different Eu concentrations to obtain more information on the formation and structure of the energy gap at the superconducting transition.

\begin{acknowledgments}
We thank Anna Pimenov and Vladimir Tsurkan for experimental support.
We are grateful to J\"{o}rg Sichelschmidt and Tobias F\"{o}rster
(Max-Planck-Institut f\"{u}r Chemische Physik fester Stoffe,
Dresden) for the opportunity to use the L-band spectrometer. We
acknowledge financial support by the Deutsche Forschungsgemeinschaft
(DFG) via the Schwerpunktprogramm SPP 1458
(Hochtemperatursupraleitung in Eisenpniktiden).  The work at POSTECH
has been supported by Basic Science Research Program (2010-0005669)
and Max Planck POSTECH/KOREA Research Initiative Program
(2011-0031558) through the National Research Foundation of Korea
(NRF).
\end{acknowledgments}

\end{document}